\documentclass[12pt]{article}
\usepackage{epsfig}
\usepackage{color}
\usepackage{amssymb,amsmath}
\usepackage{bm}
\usepackage{graphicx}

\newcommand{\bea}{\begin{eqnarray}}
\newcommand{\eea}{\end{eqnarray}}

 \makeatletter
\def\alt{\mathrel{\mathpalette\gl@align<}}
\def\agt{\mathrel{\mathpalette\gl@align>}}
\def\gl@align#1#2{\lower.6ex\vbox{\baselineskip\z@skip\lineskip\z@
\ialign{$\m@th#1\hfil##\hfil$\crcr#2\crcr\sim\crcr}}} \makeatother

\begin{document}
\begin{flushright}
KEK-TH-1327
\end{flushright}
\vspace*{1.0cm}

\begin{center}
\baselineskip 20pt 
{\Large\bf  
The minimal $B-L$ model naturally realized at TeV scale 
}
\vspace{1cm}

{\large 
Satoshi Iso$^{a,}$\footnote{satoshi.iso@kek.jp}, 
Nobuchika Okada$^{a,b,}$\footnote{okadan@post.kek.jp} 
and 
Yuta Orikasa$^{a,}$\footnote{orikasa@post.kek.jp}
} \vspace{.5cm}

{\baselineskip 20pt \it
$^{a}$ KEK Theory Center, \\
High Energy Accelerator Research Organization (KEK)  \\
and \\
Department of Particles and Nuclear Physics, \\
The Graduate University for Advanced Studies (SOKENDAI), 
\\
1-1 Oho, Tsukuba, Ibaraki 305-0801, Japan \\
\vspace{3mm}
$^{b}$Department of Physics and Astronomy, \\
University of Alabama, \\
Tuscaloosa, AL 35487, USA
} 

\vspace{.5cm}

\vspace{1.5cm} {\bf Abstract} \\
\end{center}
In a previous paper \cite{IOO}, we have proposed the minimal $B-L$
extended standard model as a 
phenomenologically viable model that realizes the Coleman-Weinberg-type
breaking of the electroweak  symmetry.
Assuming the classical conformal invariance and  stability 
 up to the Planck scale, we will show in this paper that
the model naturally predicts  TeV scale $B-L$ breaking as well as
a light standard-model singlet Higgs  boson and light right-handed neutrinos
around the same energy scale.
We also study phenomenology and detectability of the model 
at the Large Hadron Collider (LHC) and 
the International Linear Collider (ILC).

\thispagestyle{empty}

\newpage

\addtocounter{page}{-1}
\setcounter{footnote}{0}
\baselineskip 18pt
\section{Introduction}
To understand the dynamics of the electroweak symmetry breaking
is one of the most important issues in particle physics. 
In particular, the hierarchy problem, i.e. the stability of
the electroweak scale against a higher energy scale 
(e.g. GUT scale or Planck scale) is 
the most mysterious. Low energy supersymmetry provides a natural
solution, and predicts new particles around the TeV scale. 
It also predicts a relatively light Higgs boson mass below 
130 GeV compared to the standard model (SM) theoretical bound  
130 GeV$\lesssim m_h \lesssim$ 170 GeV imposed 
by the triviality and the stability of the electroweak vacuum.

We should, however, also be prepared for a case of a heavier Higgs boson 
and no signals of  supersymmetry at the experiment. 
In this case, there may be various possibilities, but
here we pay a special attention to the (almost) classical conformal 
invariance of the SM. 
Because of the chiral nature, the SM Lagrangian 
at the classical level cannot possess dimensionful parameters 
except for the Higgs mass term closely related to the gauge 
hierarchy problem.  

A common wisdom is that, even if the SM Lagrangian possesses 
the classical conformal invariance, the Higgs mass term is radiatively 
induced by matter fields with quadratically divergent coefficients,
and hence we cannot be free from the gauge hierarchy problem. 
Bardeen has argued \cite{Bardeen} that once the classical conformal invariance
and its minimal violation by quantum anomalies are imposed
on the SM, it may be free from the quadratic divergences and hence 
the gauge hierarchy problem. 
It seems difficult to realize such a mechanism in ordinary
field theories based on the Wilsonian renormalization group,
but we cannot either deny a possibility of an yet unknown
mechanism to forbid the quadratic (and possibly the quartic)
divergences in field theories based on the Planck scale physics
(see e.g. \cite{MNgravity}).
Such a mechanism inevitably requires 
 the absence of intermediate mass scales 
between the electroweak and the Planck scales.
In other words, physics at the Planck scale is
directly connected with the electroweak physics.

In this paper, 
we do not further discuss the mechanism itself, but investigate
its phenomenological implications. 
If the quadratic divergences are absent in classically
conformal theories, the conformal
symmetry is broken only by the logarithmic running of the 
coupling constants.
As a result, the electroweak symmetry breaking 
is realized not by the negative mass squared term of 
the Higgs doublet, but the radiative breaking 
a la Coleman-Weinberg (CW) \cite{CW}. 
It is, however, well-recognized that the CW scenario is already excluded
for the SM  because of the large top-Yukawa coupling.
In the original paper \cite{CW} by Coleman and Weinberg, they predicted 
the Higgs boson mass at 10 GeV assuming a small top-quark mass, 
but at present, the heavy top-quark is known to destabilize 
the Higgs potential, and the CW mechanism does not work 
(see, e.g., \cite{IOO}).
Hence we should extend the SM so that the CW mechanism works with 
phenomenologically viable parameters. 
Along this philosophy, Meissner and Nicolai \cite{MaNi}
investigated  extentions of the SM with 
the classical conformal invariance 
(see also earlier works \cite{Dias,Hempfling,Wu,Foot}).

In a previous paper \cite{IOO}, inspired by the work \cite{MaNi},
we have proposed a minimal 
phenomenologically viable model that the electroweak symmetry 
can be radiatively broken.
It is the minimal $B-L$ model \cite{B-L,B-L2}, i.e.
a $B-L$ (baryon number minus lepton number) gauged extension 
of the SM with the right-handed neutrinos and 
a SM singlet scalar field ($\Phi$) with two units of $B-L$ charge. 
The model is similar to the one proposed by Meissner and Nicolai \cite{MaNi},
but the difference is whether the $B-L$ symmetry is gauged or not.
In \cite{IOO} we showed that the gauging of $B-L$ symmetry 
plays the important role to achieve the radiative $B-L$ symmetry breaking. 
Without gauging, the Renormalization Group (RG) improved effective 
potential of $\Phi$ does not have a minimum 
(see discussions after Eq.~(\ref{phimass})). 
It is also phenomenologically favorable.

Such a model is strongly constrained by the following theoretical
requirements:
\begin{itemize}
\item Classical conformal invariance
\item Stability of the Higgs potential up to  the Planck scale
\item No other intermediate mass scales 
\end{itemize}
The electroweak as well as the $B-L$ symmetries should be
broken radiatively by the CW mechanism because of the classical
conformal invariance.
The condition that the theory is stable up to the Planck scale 
gives a strong constraint on the parameter space of the model. 
The stability of the electroweak scale against radiative corrections
gives  upper bounds for the masses of the $B-L$ gauge boson and 
the right-handed neutrinos, and in this way we are led to the minimal
$B-L$ gauged model {\it naturally realized at the TeV scale}.

In this paper, we further study the theoretical and phenomenological 
properties of the model. We first summarize the predictions of 
our model. In addition to the SM particles, the model consists of the following
new particles:
\begin{itemize}
\item gauge boson $Z'$ associated with the $B-L$ gauge symmetry
\item right-handed neutrinos $\nu_R^i$  
\item SM singlet Higgs $\Phi$ which breaks the $B-L$ gauge symmetry 
and gives the right-handed neutrinos masses
\end{itemize}
Because of the theoretical requirements, we have various 
predictions for these particles. 
The most important prediction is drawn in Fig.~\ref{FigRegion}. 
The figure shows an allowed region of the $Z'$ boson mass ($m_{Z'}$)  
and the $B-L$ gauge coupling ($\alpha_{B-L}$).
If the value of the $B-L$ gauge coupling is around the same order 
as those of the SM gauge couplings ($\alpha_{B-L} \sim 0.01$), 
$Z'$ gauge boson mass is predicted to be around a few TeV,
\begin{equation}
m_{Z'} \sim \mbox{a few TeV}
\end{equation}
and will be soon discovered at the LHC.

Another important prediction is drawn in Fig.~\ref{fig2}.
The figure shows a ratio of $m_\phi$ and $m_{Z'}$
as a function of the Yukawa coupling of the right-handed neutrino
divided by $\alpha_{B-L}$. 
The scalar mass $m_\phi$ is much smaller than the $Z'$ mass.
This is a general consequence of the CW type 
symmetry breaking where the potential minimal is realized 
by the balance between the tree level quartic Higgs potential and the
1-loop potential.
If the Yukawa couplings of the right-handed neutrinos are negligible, 
there is a simple relation for the mass ratio, 
\begin{equation}
\left( \frac{m_\phi}{m_{Z'}} \right)^2 
 \simeq \frac{6}{\pi} \alpha_{B-L} \ll 1.   
\end{equation}
If the Yukawa couplings of the right-handed neutrinos 
are larger, the scalar mass $m_\phi$ becomes much smaller.
The figure is drawn for the largest possible value of 
$\alpha_{B-L} \sim 0.01$. 
The mass ratio becomes smaller for a smaller value of
$\alpha_{B-L}$ or with the effect 
of the right-handed neutrino Yukawa coupling.
Hence $m_\phi$ is always lighter than $0.14 \ m_{Z'}$.

The figure also shows that, when the Majorana mass of 
the right-handed neutrino becomes larger than a critical value, 
it destabilizes the vacuum. 
Hence our model gives an upper bound for 
the Majorana mass of the right-handed neutrinos
$m_N^2 \lesssim 2.5 m_{Z'}^2$. 
At the maximum value of $m_N$, the SM singlet Higgs boson $\phi$ 
becomes almost massless. 
Hence the right-handed neutrinos are expected to be
similar to or 
lighter than $Z'$ gauge boson.
The lightness of the right-handed neutrinos is another
prediction of the model.

The paper is organized as follows.
In section 2, we first introduce our model and review
the analysis of the symmetry breaking studied in the
previous paper.
In section 3, we discuss the theoretical constraints
on the masses of the new particles. 
In section 4, 
we study the phenomenology of the model.
Since all the new particles should be around the TeV scale, 
a rich phenomenology can be expected at future collider experiments. 
We first study the physics of the $Z'$ gauge boson, 
which can be easily detected. 
Once $Z'$ gauge boson is found, 
it is a portal to the $B-L$ breaking sector
and to the right-handed neutrino sector.
The singlet Higgs boson can be produced associated with the $Z'$ boson 
in the same manner as the SM Higgs boson production associated with 
the $Z$ boson.  
If kinematically allowed, $Z'$ boson can decay into a pair of 
right-handed neutrinos and the nature of the seesaw mechanism 
can be revealed through this decay mode.

\section{Radiative symmetry breakings}
\subsection{Classically conformal $B-L$  model}
We first review our model.
It is the minimal $B-L$ extension of the SM \cite{B-L} 
with the classical conformal symmetry, 
and  based on the gauge group 
SU(3)$_c \times$SU(2)$_L\times$U(1)$_Y\times$U(1)$_{B-L}$. 
The particle contents (except for the gauge bosons)
are listed in Table~1 \cite{B-L2}. 
Here, three generations of right-handed neutrinos ($\nu_R^i$)
are necessarily introduced to make the model free from all 
the gauge and gravitational anomalies. 
The SM singlet scalar field ($\Phi$) works to break the U(1)$_{B-L}$ 
gauge symmetry by its VEV and at the same time, generates 
the right-handed neutrino masses. 
\begin{table}[t]
\begin{center}
\begin{tabular}{c|ccc|c}
            & SU(3)$_c$ & SU(2)$_L$ & U(1)$_Y$ & U(1)$_{B-L}$  \\
\hline
$ q_L^i $    & {\bf 3}   & {\bf 2}& $+1/6$ & $+1/3$  \\ 
$ u_R^i $    & {\bf 3} & {\bf 1}& $+2/3$ & $+1/3$  \\ 
$ d_R^i $    & {\bf 3} & {\bf 1}& $-1/3$ & $+1/3$  \\ 
\hline
$ \ell^i_L$    & {\bf 1} & {\bf 2}& $-1/2$ & $-1$  \\ 
$ \nu_R^i$   & {\bf 1} & {\bf 1}& $ 0$   & $-1$  \\ 
$ e_R^i  $   & {\bf 1} & {\bf 1}& $-1$   & $-1$  \\ 
\hline 
$ H$         & {\bf 1} & {\bf 2}& $-1/2$  &  $ 0$  \\ 
$ \Phi$      & {\bf 1} & {\bf 1}& $  0$  &  $+2$  \\ 
\end{tabular}
\end{center}
\caption{
Particle contents. 
In addition to the SM particle contents, 
the right-handed neutrino $\nu_R^i$ 
($i=1,2,3$ denotes the generation index) 
and a complex scalar $\Phi$ are introduced. 
}
\end{table}

The Lagrangian relevant for the seesaw mechanism is given as 
\bea 
 {\cal L} \supset -Y_D^{ij} \overline{\nu_R^i} H^\dagger \ell_L^j  
- \frac{1}{2} Y_N^i \Phi \overline{\nu_R^{i c}} \nu_R^i 
+{\rm h.c.},  
\label{Yukawa}
\eea
where the first term gives the Dirac neutrino mass term 
after the electroweak symmetry breaking, 
while the right-handed neutrino Majorana mass term 
is generated through the second term associated with 
the $B-L$ gauge symmetry breaking. 
Without loss of generality, we here work on the basis
where the second term is diagonalized and 
$Y_N^i$ is real and positive.

Under the hypothesis of the classical conformal invariance of the model, 
the classical scalar potential is described as 
\bea 
 V = \lambda_H (H^\dagger H)^2 + \lambda (\Phi^\dagger \Phi)^2 
   + \lambda^\prime (\Phi^\dagger \Phi) (H^\dagger H).  
\label{potential}
\eea
Note that when $\lambda^\prime$ is negligibly small, 
the SM Higgs sector and the $\Phi$ sector relevant 
for the $B-L$ symmetry breaking are approximately decoupled. 
If this is the case, we can separately analyze these two Higgs sectors. 
When the Yukawa coupling $Y_N$ is negligible compared to 
the U(1)$_{B-L}$ gauge coupling, the $\Phi$ sector is 
the same as the original Coleman-Weinberg model \cite{CW}, 
so that the radiative U(1)$_{B-L}$ symmetry breaking will be achieved. 
Once $\Phi$ develops its VEV, the tree-level mass term for the SM Higgs 
doublet is effectively generated through the third term in Eq.~(\ref{potential}). 
Taking $\lambda^\prime$ negative, the induced mass squared 
is negative and as a result, the electroweak symmetry breaking 
is driven in the same way as in the SM. 

Because of the requirement of the classical conformal invariance,
the model is characterized by a very few parameters, i.e.
besides the SM couplings,
the Dirac and Majorana Yukawa couplings for neutrinos, 
the model has only the following 4 additional parameters:
\begin{enumerate}
\item  $B-L$ gauge coupling ($\alpha_{B-L}$)
\item Higgs quartic coupling ($\lambda_H$)
\item  SM singlet Higgs quartic coupling ($ \lambda $)
\item  Mixing between $\Phi$ and $H$  ($\lambda'$).
\end{enumerate}
These four parameters determine the $B-L$ breaking scale $M$,
the electroweak breaking scale $v=246$ GeV, $m_{Z'}$, $m_\phi$
and the Higgs boson mass $m_H$. 
There is a relation between $M$, $m_{Z'}$ and $m_{\phi}$
because of the absence of the tree level mass term for the
$\Phi$ field.
\subsection{$B-L$ Symmetry Breaking} 
Without the mass terms in the scalar potential, the symmetry
breaking  must occur radiatively.  
Generally speaking, we need to study the full effective
potential for the two  Higgs fields $H$ and $\Phi$ \cite{GilWein}. 
However, since the $B-L$ breaking scale $M$ must be phenomenologically 
higher than the electroweak scale $v$ (at least one-order of magnitude), 
we can separately analyze the $B-L$ and the electroweak symmetry breakings.
In other words, since we should have a relation 
$\langle \Phi \rangle \gg \langle H \rangle$,
we can first neglect the $H$ field and calculate the 
Coleman-Weinberg potential along the $\Phi$ direction.
Then we can investigate the radiative breaking of the electroweak symmetry.
The mixing term with a negative coupling,
$\lambda^\prime (\Phi^\dagger \Phi) (H^\dagger H)$,
triggers the electroweak symmetry breaking.
The validity of this approximation can be justified for the phenomenologically
favorable parameters \cite{IOO}.

Let us first investigate the radiative $B-L$ symmetry breaking. 
We renormalize the CW effective potential at the one-loop
level as \cite{Sher, MaNi2} 
\bea 
 V (\phi) = \frac{1}{4} \lambda(t) G^4(t) \phi^4, 
\eea
where $\phi/\sqrt{2} =\Re[\Phi]$, $t=\log[\phi/M]$ 
with the renormalization point $M$, 
and 
\bea 
 G(t) = \exp \left[- \int_0^t d t^\prime \; 
 \gamma(t^\prime) \right] .
\eea 
The anomalous dimension (in the Landau gauge) is
given by 
\bea
 \gamma = \frac{1}{32 \pi^2} 
 \left[ \sum_i (Y_N^i)^2 - a_2 g_{B-L}^2 \right].  
\eea
Here, $g_{B-L}$ is the $B-L$ gauge coupling, and $a_2=24$. 
Renormalization group equations for coupling parameters involved 
in our analysis are listed below: 
\bea 
2 \pi \frac{d \alpha_{B-L}}{d t} &=&  b \alpha_{B-L}^2, 
 \nonumber \\
2 \pi \frac{d \alpha_\lambda}{d t}&=& 
   a_1 \alpha_\lambda^2 + 8 \pi \alpha_\lambda \gamma 
 + a_3 \alpha_{B-L}^2 -\frac{1}{2} \sum_i (\alpha_N^i)^2, 
 \nonumber \\ 
\pi \frac{d \alpha_N^i}{d t} &=&  \alpha_N^i 
\left(
 \frac{1}{2} \alpha_N^i + \frac{1}{4} \sum_j \alpha_N^j - 9 \alpha_{B-L} 
\right), 
\label{RGEs}
\eea
where 
$\alpha_{B-L} = g_{B-L}^2/(4 \pi)$, $\alpha_\lambda = \lambda/(4 \pi)$, 
$\alpha_N^i = (Y_N^i)^2/(4 \pi)$, and the coefficients 
 in the beta functions are explicitly given as 
 $b=12$, $a_1=10$ and $a_3=48$. 

\begin{figure}[ht]
\begin{center}
{\includegraphics*[width=.6\linewidth]{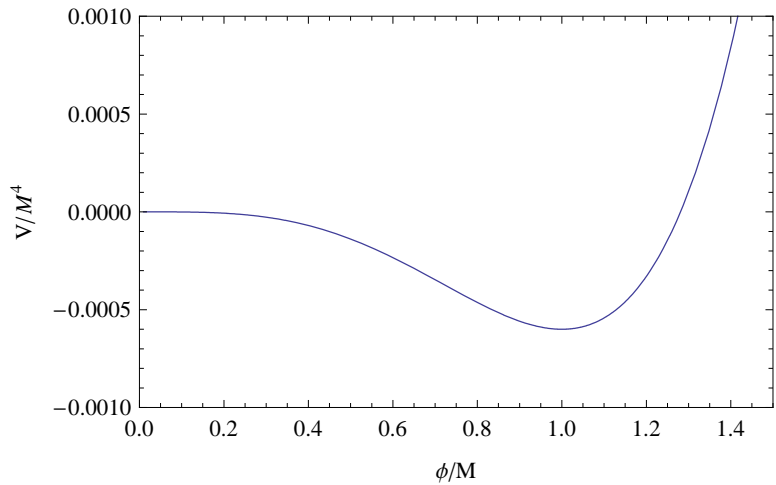}}
\caption{
The RG improved effective potential. 
Here, we have taken $\alpha_{B-L}(0)=0.01$ and 
$\alpha_N=0$ for simplicity.
}
\label{fig-CWpotential}
\end{center}
\end{figure}

By solving the RG equations, we can obtain the RG improved 
effective potential. Please refer to  \cite{IOO}
for more details of the solution.  
The Figure \ref{fig-CWpotential} depicts  the RG-improved 
 effective potential\footnote{The RG improved effective potential 
has an infrared instability, 
but it is far below the QCD scale \cite{IOO} and  
invisible in the figure.  
Since the  perturbative calculation is not reliable there,
the instability itself cannot be justifiably investigated.
Hence we do not consider it here.
} 
and it has a minimum at $\phi=M$. 

The condition for the potential to have a minimum
can be obtained without solving the RG equations.
Setting the renormalization point to be the VEV of $\phi$ 
at the potential minimum ($\phi=M$ or equivalently $t=0$), 
the stationary condition 
$
  \frac{d V}{d \phi} \Big|_{\phi=M}= 0 
$
leads to a relation among the coupling constants 
at the potential minimum (at $t=0$)
\bea 
\frac{d \alpha_\lambda}{d t} +4 \alpha_\lambda (1-\gamma) 
=\frac{1}{2 \pi}
\left(
10 \alpha_\lambda^2 + 48 \alpha_{B-L}^2 -\frac{1}{2} \sum_i
(\alpha_N^i)^2  \right) + 4 \alpha_\lambda =0  .
\label{condition}
\eea
For coupling values well within the perturbative regime, 
$\alpha_\lambda \sim \alpha_{B-L}^2 \sim (\alpha_N^i)^2  \ll 1$, 
we find the relation
\bea 
 \alpha_\lambda(0) \simeq 
 -\frac{6}{\pi} 
  \left( 
 \alpha_{B-L}(0)^2 - \frac{1}{96} \sum_i (\alpha_N^i(0))^2 
  \right) . 
\label{solution}
\eea
This is the dimensional transmutation, and 
one of the independent couplings is transmuted to the energy scale 
$M$ of the $B-L$ breaking.

The mass of the SM singlet Higgs $\phi$ can be obtained
by taking the second derivative of the effective potential
at the minimum and given by
\bea 
 m_\phi^2 = \frac{d^2 V}{d \phi^2}\Big|_{\phi=M}
  \simeq - 16 \pi \alpha_\lambda(0) M^2.
\eea
The coupling $\alpha_\lambda(0)$ satisfies the relation
Eq.~(\ref{solution}). 
We should note here that the physical coupling constant at the
minimum is given by taking the fourth derivatives of the potential as
\begin{equation}
\lambda_{eff} =  \frac{\partial^4 V}{\partial \phi^4} \Big|_{t=0}
= -\frac{22}{3} \lambda(0).
\end{equation}
Then the SM singlet Higgs boson mass is given 
in terms of the physical (effective) coupling by
\bea
m_\phi^2 &=& \frac{24 \pi}{11} \alpha_{\lambda, eff} M^2.
\label{phimass}
\eea

Therefore, 
the effective potential has a minimum at $\phi=M$ and 
the $B-L$ symmetry is radiatively broken, 
only the condition $\alpha_\lambda(0) < 0$ is satisfied. 
In the limit $\alpha_N^i \to 0$, the system is the same 
as the one originally investigated by Coleman-Weinberg \cite{CW}, 
where the U(1) gauge interaction plays the crucial role 
to achieve the radiative symmetry breaking 
keeping the validity of perturbation. 
In this sense, gauging the U(1)$_{B-L}$ is necessary 
although it is not required for the purpose to implement 
the seesaw mechanism.

\subsection{Electroweak symmetry breaking}
Now let us consider the SM Higgs sector. 
In our model, the electroweak symmetry breaking is achieved 
in a very simple way. 
Once the $B-L$ symmetry is broken, the SM Higgs doublet mass 
($\mu^2 h^2/2$)
is generated through the mixing term between $H$ and $\Phi$ 
in the scalar potential (see Eqs.~(\ref{potential})), 
\bea 
  \mu^2 = \frac{\lambda^\prime}{2} M^2.  
\label{Hmass}
\eea
Choosing $\lambda^\prime < 0$, the electroweak symmetry is 
broken in the same way as in the SM. 
However, the crucial difference from the SM is that 
in our model, the electroweak symmetry breaking originates form 
the radiative breaking of the U(1)$_{B-L}$ gauge symmetry. 
At the tree level the Higgs boson mass 
is given by $m_h^2=2 |\mu^2|=|\lambda^\prime|M^2=
2 \lambda_H v^2$
where $\langle h \rangle = v =246$ GeV. 
Then, by imposing the triviality (up to the Planck scale)
and the vacuum stability bounds, 
the Higgs boson mass is given in a range 
130 GeV$\lesssim m_h \lesssim$ 170 GeV  as in the 
SM \cite{Hmass}. 

In the case of the ordinary Coleman-Weinberg scenario for 
the SM, the large top-Yukawa coupling causes the instability
of the effective Higgs potential.
In the present case, however, the introduction of the 
$B-L$ sector saves this instability, and
the Coleman-Weinberg mechanism can be dynamically realized.
This is the theoretical reason why the $B-L$ gauge sector
is necessary  to realize the Coleman-Weinberg mechanism.

\section{Theoretical constraints on $m_{Z^\prime}, m_\phi$ and $m_N$}
Due to the theoretical requirements discussed in the introduction,
the parameter space of the model can be highly constrained.
The classical conformal invariance
reduces the number of the new coupling constants.
Then the triviality and the stability of the Higgs potential
up to the Planck scale 
strongly constrain values of the coupling constants.
Furthermore  naturalness of the electroweak scale 
against the mass scale  in the $B-L$ sector constrains 
the masses of $Z'$ and the right-handed neutrinos 
to be lighter than a few TeV. In this section we discuss these
theoretical constraints on $m_{Z^\prime}, m_\phi$ and $m_N$.

\subsection{Mass formula}
One of the two parameters in the $B-L$ sector
($\alpha_{B-L}, \ \lambda$) is used to 
determine the  $B-L$ symmetry breaking scale $M$.
Hence a relation arises between the  masses of $Z^\prime$
and $\phi$. This is due to the absence of the tree level mass term
in the classical Lagrangian of $\Phi$.
On the contrary, in the SM Higgs sector, the additional
 coupling $|H|^2 |\langle \Phi \rangle|^2$ gives  
the mass term of the Higgs doublet $H$, and then
the SM Higgs mass can be taken independently 
of the mass of the SM gauge bosons. 
 
An extra gauge boson associated with 
the U(1)$_{B-L}$ gauge symmetry acquires its mass 
through the $B-L$ symmetry breaking. 
It is given by
\bea
m_{Z'}^2 & =&16 \pi \alpha_{B-L}(0) M^2.
\eea
The coupling constant $\alpha_{B-L}$ is bounded
from above by a condition  that the running $B-L$ coupling 
does not diverge up to the Planck scale.
Roughly it is bounded as $\alpha_{B-L}(0) < 0.015$.
The constraint is drawn in Fig.~\ref{FigRegion}
as a almost straight line (in green).

On the other hand, the SM singlet Higgs mass is given
by Eq.~(\ref{phimass}) and
we can find the mass relation between 
$Z^\prime$ boson  and the SM singlet Higgs boson
\begin{equation}
\left( \frac{m_\phi}{m_{Z'}} \right)^2 
 \simeq \frac{6}{\pi} 
\left( \alpha_{B-L} - \frac{1}{96} \frac{\sum_i(\alpha_N^i)^2}{\alpha_{B-L}}
\right)  \lesssim 0.03.
\label{phimass2}
\end{equation}
The maximum value of the mass ratio is given by the maximum $\alpha_{B-L}$
and neglecting the Majorana coupling $\alpha_N$.

The hierarchy between the two masses is a general consequence 
of the Coleman-Weinberg model where the symmetry breaking occurs 
under the balance between the tree-level quartic coupling and 
the terms generated by quantum corrections. 
The scalar boson $\phi$ can be much lighter than the $Z^\prime$
gauge boson and possibly comparable with the SM Higgs boson. 
Then, as we discuss later, the two scalars  mix each other.

\begin{figure}[ht]
\begin{center}
{\includegraphics*[width=.6\linewidth]{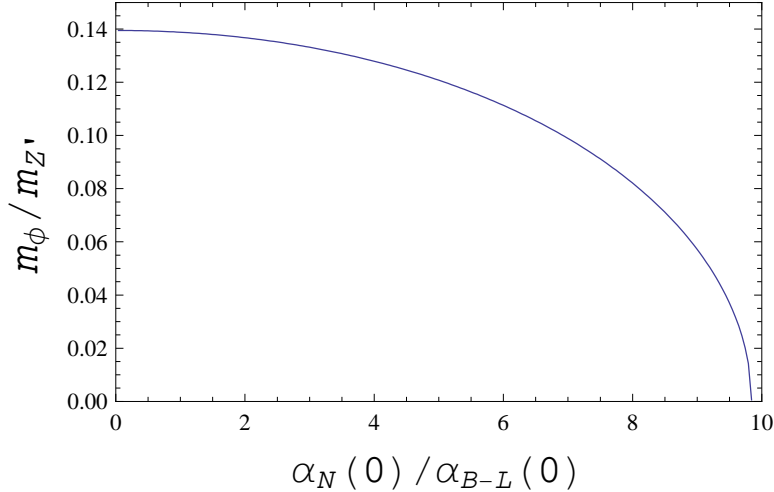}}
\caption{
The SM singlet Higgs boson mass as a function of the Yukawa coupling. 
Here we have taken $\alpha_{B-L}(0)=0.01$ and accordingly,
fixed $\alpha_\lambda(0)$ to satisfy the stationary condition 
in Eq.~(\ref{condition}). 
For $\alpha_N(0) \simeq 9.8 \alpha_{B-L}(0)$, the potential minimum 
at $\phi=M$ changes into the maximum. 
}
\label{fig2}
\end{center}
\end{figure}

 Eq.~(\ref{phimass2}) indicates that 
as the Yukawa coupling $\alpha_N$ becomes larger, 
the SM singlet Higgs boson mass squared is reducing
and eventually changes its sign. 
Therefore, there is an upper limit on the Yukawa coupling 
in order for the effective potential to have the minimum 
at $\phi=M$ ($t=0$). 
This is in fact  the same reason as  
why the Coleman-Weinberg mechanism 
in the SM Higgs sector fails to break the electroweak symmetry 
when the top-Yukawa coupling is large as observed. 
Analyzing the RG improved effective potential with 
only one Yukawa coupling $\alpha_N$, 
the SM singlet Higgs boson mass as a function of 
the Yukawa coupling is depicted in Fig.~\ref{fig2}. 
The minimum at $M$ in the effective potential changes into 
the maximum for $\alpha_N(0) > 9.8 \alpha_{B-L}(0)$.

The Majorana mass of the right-handed neutrinos must be lighter
than the critical value discussed above;
\begin{equation}
m_N^2 = Y_N^2 M^2 = 4 \pi \alpha_N M^2 <  \sqrt{6} m_{Z'}^2.
\label{RNmassbound}
\end{equation}
At the maximum value of $m_N$, the SM singlet Higgs becomes
almost massless. 
Hence the right-handed neutrinos are expected to be similar to or
lighter than $m_{Z'}$. 
This is another important prediction of our model.

\subsection{Naturalness constraints on $m_{Z'}$ and $m_N$}
We have imposed the classical conformal invariance and the absence
of the quadratic divergences to solve the gauge hierarchy problem.
This itself should be solved by the physics at the Planck scale,
but we should also take care of the loop effects of heavy states in the theory
associated with the $B-L$ breaking, since there is a small
hierarchy between the electroweak scale  $v=246$ GeV 
and the $B-L$ breaking scale $M$. 
%
Here we estimate the effects by the loop diagrams of heavy states 
on the Higgs boson mass carefully, and leads to upper bounds 
on masses of heavy states in terms of naturalness.

The states whose masses are associated with the $B-L$ breaking scale
are $Z'$ gauge boson, SM singlet Higgs boson $\phi$ 
and the right-handed neutrinos $\nu_R^i.$ 
Since the coupling $\lambda'$ between the SM singlet Higgs and 
the SM Higgs doublet is tiny, the stability of the electroweak scale 
does not give a strong constraint on the mass of $\phi$. 
Hence we will consider the effects of the right-handed neutrinos 
and the $Z'$ gauge boson.

\begin{figure}[ht]\begin{center}
\includegraphics[scale=1.2]{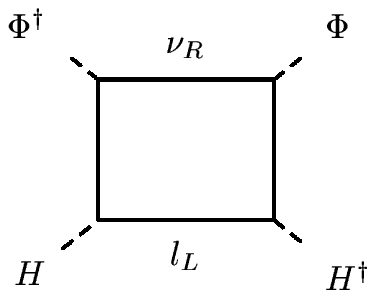}
\caption{
One-loop diagram inducing the mixing term 
 $(\Phi^\dagger \Phi)(H^\dagger H )$ 
 through the right-handed neutrinos. 
}\label{fig3}
\end{center}
\end{figure}

We first consider the one-loop effect of the right-handed neutrinos
$\nu_R^i.$ 
A typical graph with $\nu_R$ contributing to the Higgs potential 
is given by Fig.~\ref{fig3}.
When the SM singlet gets the VEV $\phi=M$, 
we obtain the effective Higgs boson mass squared such as
\begin{equation}
 \Delta m_h^2\sim \frac{Y_D^2Y_N^2}{16\pi^2}M^2
  \log\frac{M_{Pl}^2}{m_{Z'}^2}
  \sim \frac{m_\nu m_N^3}{16\pi^2 v^2}
  \log\frac{M_{Pl}^2}{m_{Z'}^2},
\end{equation}
where we have used the seesaw formula,
$m_\nu\sim Y_D^2v^2/m_N$ with $m_N=Y_N M$.
For the stability of the electroweak vacuum, 
 $\Delta m_h^2$ should be smaller than the 
electroweak scale.
(The condition is equivalent
to the {\it naturalness} of the $\lambda^\prime$ coupling.)
Thus, we can obtain the upper bound of $m_N$ once $m_\nu$ is fixed. 
For example,  when the neutrino mass is around $m_\nu \sim 0.1$ eV,
there is an upper bound for the  Majorana mass
$m_N \lesssim 2.4\times 10^6$ GeV
and hence $ M \lesssim 2.4\times 10^6/Y_N$ GeV. 
A similar constraint on the Majorana mass was found in \cite{Casas}.
In our model, the constraint is milder than Eq.~(\ref{RNmassbound}) imposed
by the stability of the potential.

\begin{figure}[ht]\begin{center}
\includegraphics[scale=1.2]{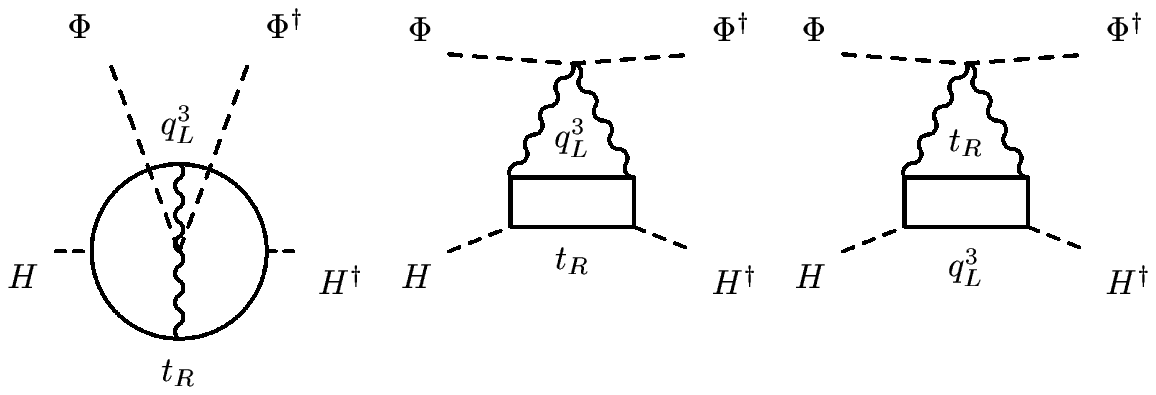}
\caption{
Two-loop diagrams inducing the mixing term $(\Phi^\dagger \Phi)(H^\dagger H )$ 
 through the top-quarks and the $B-L$ gauge bosons. 
 The wavy lines represent the 
 propagators of the $B-L$ gauge bosons.
}
\label{fig4}
\end{center}
\end{figure}

\begin{figure}[ht]\begin{center}
{\includegraphics*[width=.6\linewidth]{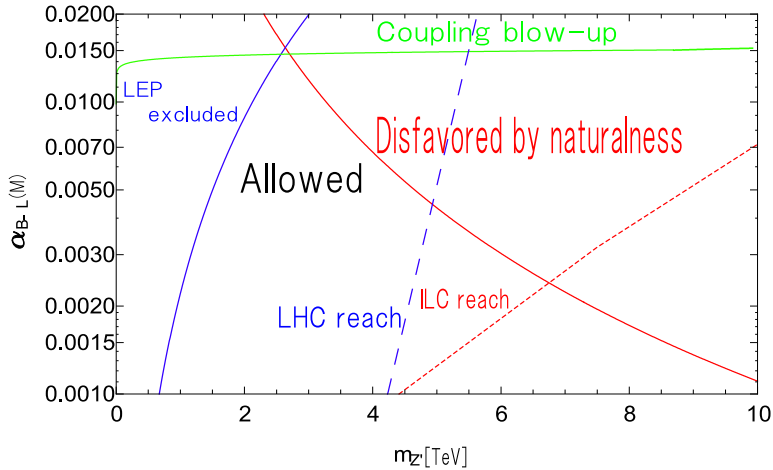}}
\caption{
The allowed parameter region is drawn.
The upper region of the almost straight line (in green) 
is rejected by a requirement that the $B-L$ gauge coupling 
does not diverge up to the Planck scale.
The upper-right side of the solid line (in red)
is disfavored by the naturalness condition of the electroweak scale.
The left of the solid line (in blue) has been already excluded
by the LEP experiment, $M \gtrsim 3$ TeV. 
The left of the dashed line can be explored in 5-$\sigma$ significance 
at the LHC with $\sqrt{s}$=14 TeV and an integrated luminosity 100 fb$^{-1}$.  
The left of the dotted line can be explored at the ILC 
with $\sqrt{s}$=1 TeV, assuming 1\% accuracy. 
}
\label{FigRegion}
\end{center}
\end{figure}

The second, but more important constraint comes from  two-loop
effects (see Fig.~\ref{fig4}) 
involving the top-quarks and the $Z'$ gauge boson.
Because of the large top-quark Yukawa coupling, these
diagrams give significant contributions.
The detail of the calculations is given in Appendix.
By substituting $\phi=M$ in the above diagrams,
 we obtain the correction  such as
\begin{eqnarray}
 \Delta m_h^2 =
 \frac{8\alpha_{B-L}m_t^2 m_{Z^\prime}^2}{\left( 4\pi\right)^3}
 \log\frac{M_{Pl}^2}{m_{Z^\prime}^2}.    
\end{eqnarray}
Note that the $(\log[M_{Pl}/m_{Z'}] )^2$ term vanishes,
and the Higgs mass correction depends linearly on the logarithm.
In order to assure the naturalness of the Higgs boson mass,
this correction cannot be much larger than the electroweak scale.
This gives a stringent constraint on the $Z'$ mass, 
which is drawn in Fig.~\ref{FigRegion} as a solid curve (in red).
The upper-right side of it is disfavored by the naturalness condition.

In Fig.~\ref{FigRegion},
the upper region of the straight line (in green) 
at $\alpha_{B-L}  \sim 0.015$ is rejected 
by a requirement that the $B-L$ gauge coupling 
does not diverge up to the Planck scale.
The upper-right side of the solid line (in red)
is disfavored by the naturalness condition discussed above.
The left of the solid line (in blue) has been already excluded
by the LEP experiment, $M \gtrsim 3$ TeV \cite{Zbound}, 
which is consistent with the bound from the direct search 
for $Z'$ boson at Tevatron \cite{CDF}. 
The figure indicates that if the $B-L$ gauge coupling is 
not much smaller than the SM gauge couplings, e.g.
$0.005 \lesssim \alpha_{B-L} \lesssim  0.015$,
the mass of the $Z'$ gauge boson is constrained to 
be around a few TeV. 
The left of the dashed (dotted) lines can be reached at
the LHC (ILC) experiment. 
We will discuss this search reach in the next section. 
Hence, most of the theoretically favorable region 
can be explored in the near future colliders.

\section{Phenomenology of TeV Scale $B-L$ Model}

Based on the simple assumption of classical conformal 
invariance, we have proposed a minimal phenomenologically 
viable model with an extra gauge symmetry. 
The naturalness of the SM Higgs boson mass constrains
the $B-L$ breaking scale to be around TeV and hence, 
the mass scale of new particles in the model, 
$Z^\prime$ boson, right-handed Majorana neutrinos and 
the SM singlet Higgs boson, is around TeV or smaller. 
These new particles may be discovered at future collider 
experiments such as the LHC and ILC. 
Now we study phenomenology of these new particles.

\subsection{Search for the $Z^\prime$ boson}
We first investigate the $Z^\prime$ boson production at the LHC. 
In our study, we calculate the dilepton production cross sections 
through the $Z^\prime$ boson exchange together with the SM processes 
mediated by the $Z$ boson and photon%
\footnote{
The quark pair production channel, in particular,
top-quark pair production via the $Z^\prime$ boson exchange 
is also worth investigating \cite{Arai}, 
since top-quark, which electroweakly decays before hadronization,
can be used as an ideal tool to probe
new physics beyond the Standard Model \cite{TopPhys}.
}.
The dependence of the cross section on the final state 
dilepton invariant mass $M_{ll}$ is described as
\begin{eqnarray}
 \frac{d \sigma (pp \to \ell^+ \ell^- X) }
 {d M_{ll}}
 &=&  \sum_{a, b}
 \int^1_{-1} d \cos \theta
 \int^1_ \frac{M_{ll}^2}{E_{\rm CMS}^2} dx_1
 \frac{2 M_{ll}}{x_1 E_{\rm CMS}^2}   \nonumber \\
&\times & 
 f_a(x_1, Q^2)
  f_b \left( \frac{M_{ll}^2}{x_1 E_{\rm CMS}^2}, Q^2
 \right)  \frac{d \sigma(\bar{q} q \to \ell^+ \ell^-) }
 {d \cos \theta},
\label{CrossLHC}
\end{eqnarray}
where $E_{\rm CMS} =14$ TeV is the center-of-mass energy of the LHC.
In our numerical analysis, we employ CTEQ5M~\cite{CTEQ} for the parton 
distribution functions with the factorization scale $Q= m_{Z^\prime}$. 
Formulas to calculate $d \sigma (\bar{q} q \to \ell^+ \ell^-)/d cos \theta$ 
are listed in Appendix.

Fig.~\ref{FigLHC} shows the differential cross section for $pp \to
\mu^+\mu^-$ for $m_{Z^\prime}=2.5$ TeV together with the SM cross 
section mediated by the $Z$-boson and photon. 
Here, we have used $\alpha_{B-L}=0.008$ and all three right-handed 
Majorana neutrino masses have been fixed to be 200 GeV as an example. 
The result shows a clear peak of the $Z^\prime$ resonance. 
When we choose a kinematical region for the invariant mass in the range,
$M_{Z^\prime}- 2 \Gamma_{Z^\prime} \leq M_{ll} \leq 
M_{Z^\prime} + 2 \Gamma_{Z^\prime}$ with 
$\Gamma_{Z^\prime} \simeq 53$ GeV, for example, 
$560$ signal events would be observed with an integrated luminosity 
of 100 fb$^{-1}$, while only a few evens are expected 
for the SM background. 
We can conclude that the discovery of the $Z^\prime$ boson with 
mass around a few TeV and the $B-L$ gauge coupling comparable 
to the SM gauge couplings is promising at the LHC.

\begin{figure}[ht]\begin{center}
{\includegraphics*[width=.6\linewidth]{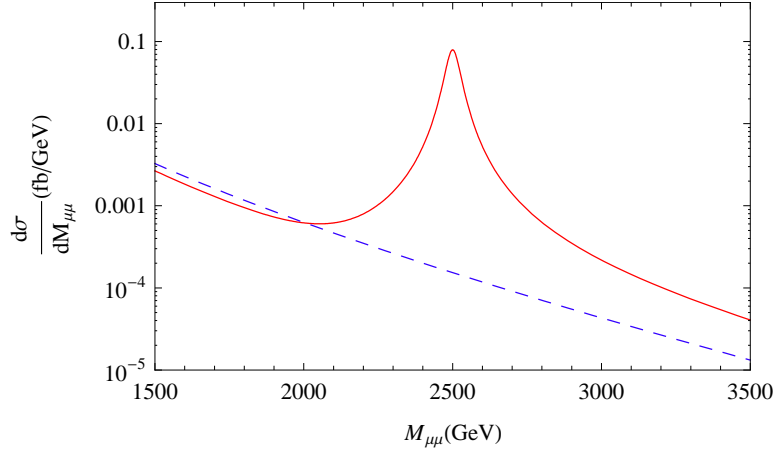}}
\caption{
The differential cross section for $pp \to \mu^+ \mu^- X$ 
at the LHC for $m_{Z^\prime}=2.5$ TeV and $\alpha_{B-L}=0.008$. 
}
\label{FigLHC}
\end{center}
\end{figure}

In order to evaluate the search reach of the $Z^\prime$ boson 
at the LHC, more elaborate study is necessary. 
We refer recent studies in \cite{Bassoetal}. 
In Fig.~\ref{FigRegion}, the dashed line (in blue) shows the 5-$\sigma$ 
discovery limit obtained in \cite{Bassoetal} for $E_{\rm CMS} =14$ TeV 
with an integrated luminosity of 100 fb$^{-1}$. 
If the $B-L$ gauge coupling is comparable to the SM ones, 
$\alpha_{B-L}={\cal O}(0.01)$, the LHC can cover the region 
$m_{Z^\prime} \lesssim 5$ TeV.

Once a resonance of the $Z^\prime$ boson has been discovered 
at the LHC, the $Z^\prime$ boson mass can be determined from 
the peak energy of the dilepton invariant mass. 
After the mass measurement, we need more precise measurement 
of the $Z^\prime$ boson properties such as couplings with 
each (chiral) SM fermion, spin and etc., in order to discriminate 
different models which predict electric-charge neutral gauge bosons. 
It is interesting to note that the ILC is capable for this task 
even if its center-of-mass energy is far below the $Z^\prime$ 
boson mass~\cite{ILC}. 
In fact, the search reach of the ILC can be beyond the LHC one.

\begin{figure}[ht]\begin{center}
{\includegraphics*[width=.6\linewidth]{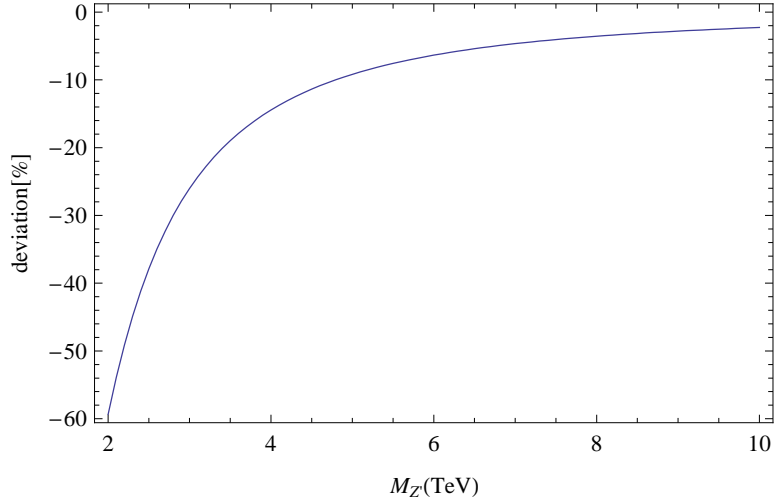}}
\caption{
Deviation (in units of \%) from the SM cross section 
as a function of $M^\prime$, 
for $\alpha_{B-L}=0.01$. 
}
\label{FigILC}
\end{center}
\end{figure}

We calculate the cross sections of the process $e^+ e^- 
\to \mu^+ \mu^-$ at the ILC with a collider energy 
$\sqrt{s}=1$ TeV for various $Z^\prime$ boson mass. 
The deviation of the cross section in our model from the SM one, 
\bea 
 \frac{\sigma(e^+ e^- \to \gamma, Z, Z^\prime \to \mu^+ \mu^-)}
{\sigma_{SM}(e^+ e^- \to \gamma, Z \to \mu^+ \mu^-)}-1,  
\eea 
is depicted in Fig.~\ref{FigILC} as a function of $m_{Z^\prime}$. 
Here we have fixed $\alpha_{B-L}=0.01$ and the differential cross 
section is integrated over a scattering angle 
$-0.95 \leq \cos \theta \leq 0.95$. 
Even for a large $Z^\prime$ boson mass, for example, 
$m_{Z^\prime}=10$ TeV, Fig.~\ref{FigILC} shows a few percent deviations, 
which is large enough for the ILC with an integrated luminosity 
500 fb$^{-1}$ to identify. 
Assuming the ILC is accessible to 1 \% deviation, 
the search limit at the ILC has been investigated in \cite{Bassoetal} 
and in Fig.~\ref{FigRegion}, the dotted line (in red) shows the result. 
The ILC search limit is beyond the one at the LHC. 

The discovery of the $Z^\prime$ boson has also an impact on neutrino physics. 
In our model, tiny neutrino masses are obtained by the seesaw mechanism 
after integrating out the heavy right-handed Majorana neutrinos. 
As we discussed in the previous section, there is an upper bound on 
the Yukawa coupling of the right-handed neutrinos 
not to destabilize 
the $B-L$ symmetry breaking minimum, $\alpha_N < 9.8 \alpha_{B-L}$. 
Thus, it is likely that right-handed neutrinos are light 
enough to be produced by $Z^\prime$ boson decay. 
If this is the case, the way to investigate the seesaw mechanism 
is opened up.  
Once the $Z^\prime$ boson is produced, it decays into a right-handed 
neutrino with a branching ratio 
${\rm Br}(Z^\prime \to \nu_R^i \nu_R^i) \simeq 6$ \% 
(see Appendix for partial decay widths of the $Z'$ boson). 
Produced right-handed neutrinos decay into the SM Higgs boson and 
the weak gauge bosons through the mixing with the SM left-handed 
neutrinos and provide a signature at the LHC 
through trilepton final states with a small SM background \cite{NuRLHC1} 
and events of like-sign leptons associated with 
the Majorana nature of the right-handed neutrinos \cite{NuRLHC2}.

The leptogenesis \cite{F-Y} through the lepton number and 
CP violating decays of the right-handed Majorana neutrino 
is a very simple mechanism for baryogenesis, 
the origin of the baryon asymmetry in the universe. 
In normal thermal leptogenesis scenario, 
there is a lower mass bound on the lightest right-handed neutrino, 
$m_N \gtrsim 10^9$ GeV \cite{LowerBound}, 
in order to achieve the realistic baryon asymmetry 
in the present universe. 
In our model, the right-handed neutrino mass is around TeV scale 
or smaller and far below this bound. 
In this case, the leptogenesis can be possible 
through the resonant leptogenesis mechanism \cite{resonantLG} 
due to an enhancement of the CP asymmetry parameter 
$\epsilon$ via well-degenerated right-handed neutrinos. 
If the CP asymmetry parameter is sufficiently large 
by the resonant leptogenesis mechanism, the difference between 
the number of positive and negative like-sign dileptons 
from the right-handed neutrino decays can be detectable 
at the LHC \cite{Blanchetetal}.

\subsection{Phenomenology of the SM singlet Higgs boson}
Let us finally consider another new particle in the present model, 
the SM singlet Higgs boson $\phi$, and its phenomenological 
implications. 
According to the analysis in the previous section, 
the singlet Higgs boson is relatively light to the $Z^\prime$ boson, 
for example, an order of magnitude lighter 
for $\alpha_{B-L} \sim 0.01$. 
This fact is a general consequence from radiative symmetry 
breaking by the CW mechanism. 
This light SM singlet Higgs can be mixed up with the SM Higgs 
boson through the third term in Eq.~(\ref{potential}) and thus, 
it potentially affects phenomenology of Higgs boson \cite{HiggsPheno}.

In our model, because of the absence of tree-level mass terms 
under the classical conformal invariance, the mixing angle 
between the SM Higgs and singlet Higgs bosons is not 
an independent parameter. 
Using the relation $\lambda^\prime = -m_h^2/M^2$, 
the scalar mass matrix is given by 
\bea 
{\cal M}=
\left(
\begin{array}{cc}
m_h^2 & -m_h^2 \left( \frac{v}{M} \right) \\
-m_h^2 \left( \frac{v}{M} \right) & m_\phi^2 
\end{array}\right),  
\end{eqnarray}
where $m_\phi^2 = 96 \alpha_{B-L}^2 M^2$ 
by neglecting $\alpha_N^i$, for simplicity. 
The mixing angle is described as 
\bea  
\tan 2 \theta = 
\frac{2 m_h^2 (v/M)}{m_h^2 - 96 \alpha_{B-L}^2 M^2}. 
\eea 
For $m_h^2 \gtrsim m_\phi^2$, $\tan 2 \theta \sim 2 v/M \lesssim 0.1 $, 
while $\tan 2 \theta \sim -2 (m_h/m_\phi)^2 (v/M) \ll 1 $ 
for $m_h^2 \ll m_\phi^2$. 
In both cases, the mixing angle is small.

In models with multiple SM singlet Higgs scalars, the restrictions 
on the parameter space from precision electroweak measurements 
have been investigated \cite{Dawson}. 
Contributions of the SM singlet Higgs boson to 
the $S$, $T$ and $U$ parameters are roughly proportional 
to the mixing angle squared \cite{Peskin} and hence 
negligible in our case. 
We have checked that the range of SM Higgs boson mass 
favored by the electroweak precision measurements is 
shifted, at most, by a few GeV in our model.

If the singlet Higgs boson is light enough, the SM Higgs boson 
can decay into a pair of the singlets. 
This partial decay width is found to be 
\bea 
\Gamma(h\rightarrow \phi\phi)=\frac{1}{32\pi}\frac{v^2m_h^3}{M^4}\sim 
  1.6\times 10^{-5} \; {\rm GeV}  
\eea 
for $m_h=130$ GeV and $M=3$ TeV, for example. 
We compare this width to the partial decay width 
into bottom quarks, 
\bea 
 \Gamma(h\rightarrow  b{\bar b})=
  \frac{3}{8\pi}\frac{m_b^2}{v^2}m_h\sim 2.3\times 10^{-3} \; {\rm GeV}, 
\eea 
which is the dominant decay mode for $m_h=130$ GeV. 
The branching fraction into a pair of the singlet Higgs bosons 
is small, ${\rm Br}(h \to \phi \phi)={\cal O}(0.01)$. 
Once the singlet Higgs bosons are produced, 
they mainly decay into bottom quarks through 
the mixing with the SM Higgs boson. 
Note that this process, Higgs boson production and its decay into 
a pair of the singlet Higgs boson followed by their decays into 
bottom quarks, is similar to the Higgs boson pair production 
through the Higgs self-coupling. 
Here, let us consider Higgs pair production associated with 
$Z$ boson, $e^+e^- \to Z^* \to Z h^* \to Z h h $, at the ILC \cite{Takubo}. 
The production cross section is of order 0.1 fb 
for a collider energy 500 GeV-1 TeV. 
On the other hand, the pair of the singlet Higgs bosons are 
produced via $e^+e^- \to Z^* \to Z h$, followed by 
the SM Higgs decay $h \to \phi \phi$. 
In fact, this production cross section is comparable to 
the one for the Higgs pair production, 
$ \sigma = \sigma(e^+ e^- \to Z h) \times {\rm Br}(h \to \phi \phi)
\sim {\cal O}(10 {\rm fb}) \times 0.01 \sim 0.1$ fb. 
Experiments for precision measurements of the Higgs self-coupling 
may reveal the light singlet Higgs boson.

Once the $Z^\prime$ boson is discovered, the best way to search 
for the singlet Higgs boson would be its production associated 
with $Z^\prime$ boson, which is analogous to 
the SM Higgs boson search at LEP2, for example. 
If a future collider such as the ILC has its energy high enough 
to produce $Z^\prime$ boson and the singlet Higgs boson, 
it would be easy to discover the singlet Higgs boson.

\section{Conclusions}
In this paper we have investigated the minimal $B-L$ model
as a phenomenologically viable model in which the Coleman-Weinberg
mechanism of the electroweak symmetry breaking does work.
Because of the theoretical requirements of the 
classical conformal invariance, stability of the potential
up to the Planck scale and the naturalness of the electroweak
scale against the $B-L$ sector, the parameter region
of the model is strongly constrained. 
We are thus led to the minimal $B-L$ model naturally realized
at the TeV scale. 
This is a remarkable finding because in the usual minimal $B-L$ model, 
there is no special reason for the $B-L$ symmetry breaking scale 
to be around the TeV scale and the breaking scale can be 
much higher without any theoretical and experimental problems.

According to the TeV scale $B-L$ symmetry breaking, 
all new particles introduced in addition to the SM particles 
have masses around the TeV scale or even smaller, 
so that some signatures of such particles can be 
expected at future experiments. 
In particular, the $Z'$ boson with a few TeV mass is promising 
to be discovered at the LHC when the $B-L$ gauge coupling 
is not too small, say, $\alpha_{B-L} \gtrsim 0.005$. 
Once the $Z'$ boson is discovered, 
it will be a portal to explore the $B-L$ sector.  
In our model, it is very likely that the right-handed neutrinos 
are sufficiently light so that it can be pair-produced by 
the $Z'$ boson decay. 
Through the $Z'$ boson production, the seesaw mechanism 
can be investigated at the future colliders.

A general consequence of the CW symmetry breaking is 
a prediction of a relatively light SM singlet Higgs boson 
compared to the $Z'$ gauge boson. 
We have shown that the SM singlet Higgs boson can be 
as light as the SM Higgs boson. 
Although this light singlet Higgs boson can mix 
with the SM Higgs boson and potentially affects 
on Higgs phenomenology, the effects are found to be small 
because of a small mixing angle according to the absence 
of the tree-level mass terms in the scalar potential. 
Precision measurements of the Higgs self-couplings 
may reveal the existence of the SM singlet Higgs boson. 
Again, once the $Z'$ boson is discovered, the $Z'$ boson 
can be used to discover the light singlet Higgs boson. 
For example, if the energy of the ILC is high enough, 
$\sqrt{s} > m_{Z'} +m_{\phi}$, 
the singlet Higgs boson can be produced through 
the process, $e^+ e^- \to Z^{\prime *} \to Z' \phi$, 
which is completely analogous to the process 
at LEP2 experiment to search for the SM Higgs boson. 
In order to confirm the CW symmetry breaking, 
it is crucial to check the mass relation in Eq.~(\ref{phimass}).

\newpage
\noindent{\Large \bf Appendix}
\appendix
\section{Calculation of the two-loop diagrams}
\begin{figure}[ht]\begin{center}
\includegraphics[scale=1.5]{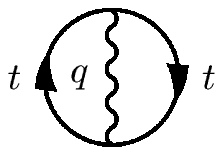}
\caption{
2-loop vacuum diagram of  a top-quark and a $Z'$ gauge boson
with momentum $q$.
}
\label{FigVacuum}
\end{center}
\end{figure}

In this appendix we calculate the two loop diagrams in Fig.~\ref{fig4}.
In order to evaluate them, we first calculate the 2-loop vacuum 
diagram of Fig.~\ref{FigVacuum}, and then take  terms proportional
to $m_{Z'}^2 m_t^2$ by taking the second derivatives with respect to $m_{Z'}$
and $m_t$.

The 2-loop vacuum diagram can be easily calculated as
\begin{eqnarray}
   \int\frac{d^d q}{\left(2\pi\right)^d}
  i\Pi_2(q^2)\left( q^2g^{\mu\nu}-q^\mu q^\nu\right)\frac{-ig^{\mu\nu}}{q^2-m_{z^\prime}^2}
\end{eqnarray}
in terms of the top-quark contribution to 
the 1-loop self-energy diagram of the $Z'$ boson
\begin{equation}
\Pi^{\mu \nu}(q) = i\left( q^2g^{\mu\nu}-q^\mu q^\nu\right) \Pi_2(q^2), 
\end{equation}
where $\Pi_2(q^2)$ is given by
\begin{eqnarray}
 \Pi_2(q^2)=-\frac{8g_{B-L}^2}{3\left(4\pi\right)^{\frac{d}{2}}}
 \int^1_0 dx x(1-x)\frac{\Gamma\left(2-\frac{d}{2}\right)}{\left( m_t^2-x(1-x)q^2\right)^{2-\frac{d}{2}}}.
\end{eqnarray}
Then by taking derivatives we can obtain 
the terms proportional to $m_{Z'}^2 m_t^2$ as
\begin{eqnarray}
 \frac{\partial^2}{\partial m_t^2\partial m_{z^\prime}^2}
 \int\frac{d^d q}{\left(2\pi\right)^d}i\Pi_2(q^2)\left(q^2g^{\mu\nu}-q^\mu q^\nu\right)\frac{-ig^{\mu\nu}}{q^2-m_{z^\prime}^2} \nonumber\\
 =-\frac{8ig_{B-L}^2\Gamma\left(3-\frac{d}{2}\right)}{3\left(4\pi\right)^
  {\frac{d}{2}}}\int\frac{d^d q}{\left(2\pi\right)^d}\frac{d-1}{q^4} 
 \sim-\frac{8ig_{B-L}^2}{\left(4\pi\right)^4}\log\frac{M_{Pl}^2}{m_{z^\prime}^2}.
\end{eqnarray}

\section{Helicity amplitudes}\label{hel-amp}
Here we provide formulas useful for calculations of 
cross sections discussed in this paper. 
We begin with the following interaction between 
a massive gauge boson ($A_\mu$) with mass $m_A$ and 
a pair of the SM fermions, 
\begin{eqnarray}
 {\cal L}_{\rm int}= J^\mu A_{\mu}
 =\bar{\psi}_f \gamma^\mu(g_L^f P_L+g_R^fP_R)\psi_f A_{\mu} .
\end{eqnarray}
A helicity amplitude for the process 
$f(\alpha)\bar{f}(\beta)\rightarrow F(\delta)\bar{F}(\gamma)$ 
is given by
\begin{eqnarray}
 {\cal M}(\alpha,\beta;\gamma,\delta)=
 \frac{g_{\mu \nu}}{s - m_A^2 + i m_A \Gamma_A}
J_{\rm in}^\mu(\alpha, \beta) J_{\rm out}^\nu(\gamma, \delta),
\end{eqnarray}
where $\alpha, \beta$ ($\gamma,\delta$) denote initial (final) 
spin states for fermion and anti-fermion, respectively, and 
$\Gamma_A $ is the total decay width of the $A$ boson. 
We have used 't Hooft-Feynman gauge for the gauge boson propagator 
and there is no contribution from Nambu-Goldstone modes 
in the process with the massless initial states.  

The currents for initial (massless) and final (massive) states 
are explicitly given by
\begin{eqnarray}
 J_{\rm in}^\mu(+,-)=-\sqrt{s}g_R^f(0,1,i,0)\,,~~~~~
 J_{\rm in}^\mu(-,+)=-\sqrt{s}g_L^f(0,1,-i,0)\,,
\end{eqnarray}
and
\begin{eqnarray}
 J_{\rm
  out}^{\mu}(+,+)&=&\omega_+\omega_-\left[g_L^F(1,-\sin\theta,0,-\cos\theta)
  -g_R^F(1,\sin\theta,0,\cos\theta)\right], 
\nonumber \\
 J_{\rm
  out}^{\mu}(-,-)&=&\omega_+\omega_-\left[g_L^F(1,\sin\theta,0,\cos\theta)
  -g_R^F(1,-\sin\theta,0,-\cos\theta)\right], 
\nonumber \\
 J_{\rm
  out}^{\mu}(+,-)&=&\omega_-^2g_L^F(0,-\cos\theta,i,\sin\theta)
  -\omega_+^2g_R^F(0,\cos\theta,-i,-\sin\theta), 
\nonumber \\
 J_{\rm
  out}^{\mu}(-,+)&=&\omega_+^2g_L^F(0,-\cos\theta,-i,\sin\theta)
  -\omega_-^2g_R^F(0,\cos\theta,i,-\sin\theta)\,,
\end{eqnarray}
where 
 $\theta$ is the scattering angle, 
 $\omega_\pm^2={\sqrt{s}\over 2}(1\pm \beta_F)$, 
 $\beta_F=\sqrt{1-{4m_F^2 \over s}}$, and 
 $f$($F$) denotes a flavor of initial (final) state of fermions.

The couplings for the SM $Z$ boson are as follows: 
\begin{eqnarray}
 g_L^\nu&=&{e \over \cos\theta_W\sin\theta_W}{1 \over 2}\,,~~~g_R^\nu=0, 
\nonumber \\
 g_L^l&=&{e \over \cos\theta_W\sin\theta_W}
   \left(-{1 \over 2}-\sin^2\theta_W(-1)\right)\,,
  ~~~g_R^l=-e(-1)\tan\theta_W, 
\nonumber \\ 
 g_L^u&=&{e \over \cos\theta_W\sin\theta_W}
   \left({1 \over 2}-\sin^2\theta_W{2 \over 3}\right)\,,
  ~~~g_R^u=-e{2 \over 3}\tan\theta_W, 
\nonumber \\ 
 g_L^d&=&{e \over \cos\theta_W\sin\theta_W}
   \left(-{1 \over 2}-\sin^2\theta_W\left(-{1 \over 3}\right)\right),
  ~~~g_R^d=-e\left(-{1 \over 3}\right)\tan\theta_W, 
\end{eqnarray}
where $\theta_W$ is the weak mixing angle. 
The couplings for the $Z^\prime$ boson are much simpler: 
\begin{eqnarray}
&&  g_L^\nu = g_R^\nu = g_L^l = g_R^l = -1, \nonumber \\
&&  g_L^u   = g_R^u   = g_L^d = g_R^d = \frac{1}{3}.
\end{eqnarray}

\section{Decay width}\label{appendix-C}
Explicit formulas of the partial decay widths of $Z^\prime$ boson 
are the following: 
\begin{eqnarray}
&& \Gamma(Z' \rightarrow \nu_l^i \bar{\nu_l^i}) 
 = \frac{m_{Z'}}{24 \pi} , 
\nonumber \\
&& \Gamma(Z' \rightarrow \nu_h^i \bar{\nu_h^i}) 
 = \frac{m_{Z'}}{24 \pi} \left( 1- \frac{m_N^{i 2}}{m_{Z'}^2} \right) 
   \sqrt{1- 4 \frac{m_N^{i2}}{m_{Z'}^2}} ,  
\nonumber \\
&& \Gamma(Z' \rightarrow e^+ e^-/\mu^+ \mu^-/\tau^+ \tau^-) 
 = \frac{m_{Z'}}{12 \pi}  ,
\nonumber \\
&& \Gamma(Z' \rightarrow u \bar{u}/c \bar{c}) 
 = \Gamma(Z' \rightarrow d \bar{d}/s \bar{s}/b \bar{b} ) 
 = \frac{m_{Z'}}{36 \pi} , 
\nonumber \\
&& \Gamma(Z' \rightarrow t\bar{t})
 = \frac{m_{Z'}}{36 \pi} 
  \left( 1- 2 \frac{m_t^2}{m_{Z'}^2} \right)  
  \sqrt{ 1- 4 \frac{m_t^2}{m_{Z'}^2}} ,
\end{eqnarray}
where $i=1,2,3$ is the generation index, and  
$\nu_l$ ($\nu_h$) denotes the light (heavy) Majorana 
neutrino mass eigenstate after the seesaw mechanism.

\section*{Acknowledgments}
We would like to thank Toru Goto
for useful discussions on the 2-loop calculations. 
The work of N.O. is supported in part by the Grant-in-Aid 
for Scientific Research from the Ministry of Education, 
Science and Culture of Japan, No.~18740170.



\end{document}